\documentclass{iaus}
\usepackage{graphicx}

\title[Methanol Masers and Star Formation] %% give here short title %%
{Methanol Masers and Star Formation}

\author[Sobolev, Ostrovskii, Kirsanova, Shelemei, Voronkov, Malyshev]   %% give here short author list %%
{A.M. Sobolev$^1$, A.B. Ostrovskii$^1$, M.S. Kirsanova$^1$,\break
O.V. Shelemei$^1$, M.A. Voronkov$^2$ \and A.V. Malyshev$^1$}

\affiliation{$^1$Ural State University, Ekaterinburg, 620083,
Russia \break
email: Andrej.Sobolev@usu.ru, Andrei.Ostrovskii@usu.ru\\[\affilskip]
$^2$ATNF CSIRO, Sydney, Australia \break email:
Maxim.Voronkov@csiro.au}

\pubyear{2005}
\volume{227}  %% insert here IAU Symposium No.
\pagerange{??--??}
\date{?? and in revised form ??}
 \jname{Massive Star Birth: A Crossroads of
Astrophysics} \editors{E.Churchwell, P.Conti \& M.Felli, eds.}
\begin{document}

\maketitle

\begin{abstract}
Methanol masers which are traditionally divided into two classes
provide possibility to study important parts of the star forming
regions: Class~II masers trace vicinities of the massive YSOs
while class~I masers are likely to trace more distant parts of the
outflows where newer stars can form.

There are many methanol transitions which produce observed masers.
This allows to use pumping analysis for estimation of the physical
parameters in the maser formation regions and its environment, for
the study of their evolution. Extensive surveys in different
masing transitions allow to conclude on the values of the
temperatures, densities, dust properties, etc. in the bulk of
masing regions. Variability of the brightest masers is monitored
during several years. In some cases it is probably caused by the
changes of the dust temperature which follow variations in the
brightness of the central YSO reflecting the character of the
accretion process.

A unified catalogue of the class II methanol masers consisting of
more than 500 objects is compiled. Analysis of the data shows
that: physical conditions within the usual maser source vary
considerably; maser brightness is determined by parameters of some
distinguished part of the object - maser formation region; class
II methanol masers are formed not within the outflows but in the
regions affected by their propagation.

It is shown that the "near" solutions for the kinematic distances
to the sources can be used for statistical analysis. The
luminosity function of the 6.7~GHz methanol masers is constructed.
It is shown that improvement of the sensitivity of surveys can
increase number of detected maser sources considerably.

The distribution of class II methanol masers in the Galaxy is
constructed on the basis of estimated kinematic distances. It is
shown that most of the sources are located in the Molecular Ring
and that the dependence of the number of sources on the distance
from the Galactic Center has significant peaks at the positions
corresponding to the spiral arms.

A survey of CS(2-1) line emission tracing dense gas is performed
at Mopra toward the positions of the brightest class II methanol
masers. Velocity correlations between the maser and CS lines are
analyzed. It is shown that the sources with l from 320 to 350 deg
in which the masers are relatively blue-shifted, form a group
which is located in the region of the Scutum-Centaurus spiral arm.
This can reflect existence of a grand design, i.e., grouping of
the sources with similar peculiarity of morphology or evolutionary
stage of the massive star forming regions.

%% add here a maximum of 10 keywords, to be taken form the file <Keywords.txt>
\end{abstract}

\firstsection % if your document starts with a section,
              % remove some space above using this command.
\section{Introduction}

Methanol masers are traditionally divided into two classes.
Although masers of both classes often co-exist in the same
star-forming region, they are usually seen apart from each other.
Class~II masers are found in the vicinity of the massive YSOs
while class~I masers are believed to trace distant parts of the
outflows from these YSOs.

Class II methanol masers display strong emission in transitions at
6.6 and 12.1 GHz (\cite{men91}). More than 2 dozens of weaker
class II maser transitions are detected in several sources (see,
e.g., \cite{sob02}). At present there exist 3 unified catalogues
of class II methanol masers containing more than 500 objects
(\cite{mal03}, \cite{xu03} and \cite{pest05}) and references to
the initial data. Distinguishing feature of the \cite{mal03}
catalogue is that it contains data on several maser transitions
and cross-references to the data on molecular shock tracers. Most
of the class II maser sources were found toward positions of the
IRAS sources while the blind surveys brought some detections in
the places without other evidences of star formation. At present
class II masers were found only in the high mass star forming
regions (\cite{min03}).

Class I methanol masers are less strong and widespread. Most of
them were found in vicinities of class II maser sources (see,
e.g., \cite{sly94} and \cite{ell05}). There were extensive
searches (e.g., \cite{sly94} and \cite{val00}) for class I sources
but no unified catalogues of these objects do exist and no blind
surveys were performed yet.

\section{Determination of parameters of the high mass star forming regions}\label{sec:param}

\subsection{Class I methanol masers}
The transitions of the \hbox{J$_2$ -- J$_1$E} series at about
25~GHz in OMC-1 were the first methanol masers detected in the
interstellar medium. Existence of the interferometry data
(\cite{joh92}) allowed construction of the models constraining
characteristics of the turbulence in this source (\cite{sww,swo}).

Analysis of existing observational data on class I methanol masers
allows to distinguish 4 regimes differing by the series of the
brightest (in terms of brightness temperature) line.

Most widespread maser regime is likely to have brightest lines
which belong to the \hbox{J$_{-1}$ -- (J-1)$_0$E} series. The
lines \hbox{4$_{-1}$ -- 3$_0$E} at 36.1~GHz and \hbox{5$_{-1}$ --
4$_0$E} at 84.5~GHz are likely to be a weak masers under normal
conditions of the massive star forming region. Usually maser
nature of the lines in this regime is difficult to prove
observationally. Anyhow, there are cases when the line profiles
contain narrow spikes and the maser nature is proved
interferometrically. The sources Sgr B2, G30.8-0.1, and G1.6-0.025
can be considered as representatives of this maser regime
(\cite{sob96}).

In the other maser regime the lines of the \hbox{J$_0$ --
(J-1)$_1$A$^+$} series become prevalent. Numerous sources manifest
definitely maser lines arising in the \hbox{7$_0$ -- 6$_1$A$^+$}
and \hbox{J$_0$ -- (J-1)$_1$A$^+$} transitions at 44.1 and
95.2~GHz, respectively. Masers in the sources DR21W, NGC2264 and
OMC-2 represent this regime (\cite{men91}). Preliminary
theoretical analysis of the pumping shows that the lines of
\hbox{J$_0$ -- (J-1)$_1$A$^+$} series become brightest in the
models with rather high beaming ($> 20$) and moderate column
densities (Sobolev, Ostrovskii \& Voronkov, in prep., presented in
the poster at IAUS227).

Next maser regime is less widespread and is represented by the
sources where the lines of the \hbox{J$_2$ -- J$_1$E} series at
about 25~GHz are the brightest. OMC-1 is the famous example of
such sources (\cite{joh92}) and some bright sources were found
recently (\cite{vor05}). Lines of this series become brightest in
the models with the high specific column densities
($lg(N_{CH_3OH}/\Delta V$~[cm$^{-3}\cdot$c])$ \ge 12$) and require
relatively high temperatures ($ T_k= 75-100$~K) and densities
($lg(n_H$~[cm$^{-3}])=5-7$).

\begin{figure}
\vspace{0cm}
\begin{center}
\includegraphics[width=5.3in,angle=0]{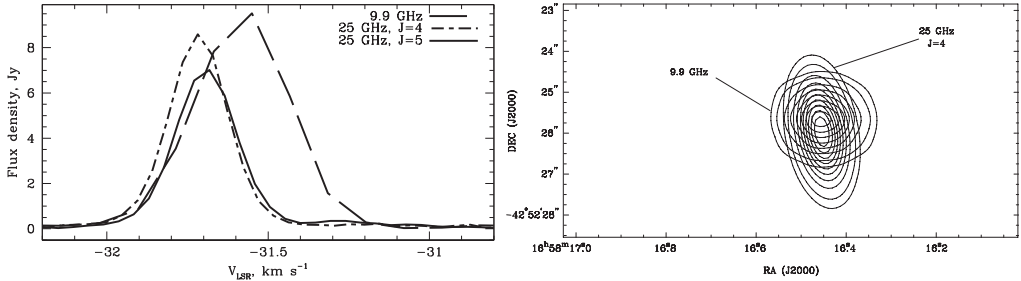}
\caption{Spectra and maps of the class I methanol masers at
9.9~GHz and $\sim$25~GHz in G343.12-0.06. Contours in the map are
(1,2,3,4,5,6,7,8,9)x0.9 Jy/beam.}\label{fig:fg1}
\end{center}
\end{figure}

Existence of the forth maser regime with the brightest line at
9.9~GHz from the \hbox{J$_{-2}$ -- (J-1)$_{-1}$E} series was
previously uncertain. Recent ATCA observations have shown that
this regime is actually realized in the sources W33A and
G343.12-0.06 (see figure~\ref{fig:fg1}). Preliminary modelling
shows that the \hbox{ 9$_{-2}$ -- 8$_{-1}$E} line at 9.9~GHz
becomes brightest in the models with specific column densities
which are as high as required for the previous regime but prefers
greater beaming and either lower densities or higher temperatures
($ T_k > 100$~K).

So, class I methanol masers are potentially a good probe of such
physical parameters as the temperature, density and column
density.

\subsection{Class II methanol masers}
Both observations and modelling of class II masers show that the
6.6~GHz transition always manifests the highest brightness
temperature (\cite{mal03}, \cite{sob97}). Observations did not
find significant differences in positions of the maser spots (see,
e.g., \cite{men92}, \cite{sut01}). So, there is basically only one
known pumping regime of class II masers. However, ratios of the
brightnesses of different maser lines show considerable dependence
on the values of the physical parameters. This allows
determination of the physical parameters in individual sources (on
the basis of multi-transitional observations, e.g., \cite{sut01},
\cite{cra01}) and construction of the model of the "common" class
II methanol maser source (on the basis of extensive surveys, e.g.,
\cite{cra04} and \cite{ell04}). Results of such studies are
summarized in \cite{sob02} and \cite{cra05} and will not be
considered here.

Different response of the maser fluxes to the changes of the
physical parameters provides possibility to elucidate nature of
the time variability and estimate physical parameters in the
source via modelling of variability in the different maser lines.
For example, interferometric studies of variability in the source
G9.62+0.20 show that the changes in the different parts of the
source are synchronized with the speed of light (\cite{goe05}). In
this case the variability is most probably caused by the changes
of the dust temperature which follow variations in the brightness
of the central YSO reflecting the character of the accretion
process.

\section{Some topics of the methanol maser statistics}\label{sec:stat}

\subsection{Distribution of the methanol masers in the Galaxy}
Extensive surveys have shown that the radial velocities of class
II methanol masers are usually coincident with those of associated
molecular cores to within several km/s (see, e.g.,
\cite{mal03,sly99} and \cite{szy05}). This provides a possibility
for making estimates of distances to these sources using models of
rotation curves of the gas in the Galaxy (the kinematic
distances). On the basis of such estimates several authors
discussed the distribution of methanol maser sources in the Galaxy
(see, e.g.,\cite{vdw97,sly99} and \cite{pest05}). It should be
noted that for the most of the sources kinematic estimates are
ambiguous and allow 2 solutions: "near" and "far" ones. Anyhow, it
was firmly established that the most of methanol masers are
situated in the Molecular Ring of the Galaxy.

We addressed the question of the kinematic distance ambiguity from
the statistical point of view. Of course, sometimes the "near"
estimate is closer to the real distance to the source and
sometimes the "far" estimate appears to be more realistic. We
looked at the extremes. This is reasonable because in the top view
of the maser distribution there is a pronounced dip of the maser
number density between the "near" and "far" distance domains.

Firstly, we assumed that the "far" estimate of kinematic distance
is correct. It was found that in this case the dependence of the
average peak maser flux on the distance from the Sun is odd and
shows general tendency to increase when the distances exceed 5
kpc. Such dependence is certainly unrealistic.

Secondly, we considered the dependence of the average peak maser
flux on the distance from the Sun for the "near" distance
estimates. In this case there is a pronounced tendency of the
decrease with the distance from the Sun. Taking into account that
the weaker masers at greater distances are missed due to the
limited sensitivity of the surveys we obtained the dependence
which is quite close to the natural, i.e., fall which is
proportional to the square of the distance from the Sun. So,
confinement to the "near" kinematic distances is likely to produce
statistical results close to reality.

\begin{figure}
\vspace{0cm}
\begin{center}
\includegraphics[width=2.45in,angle=-90]{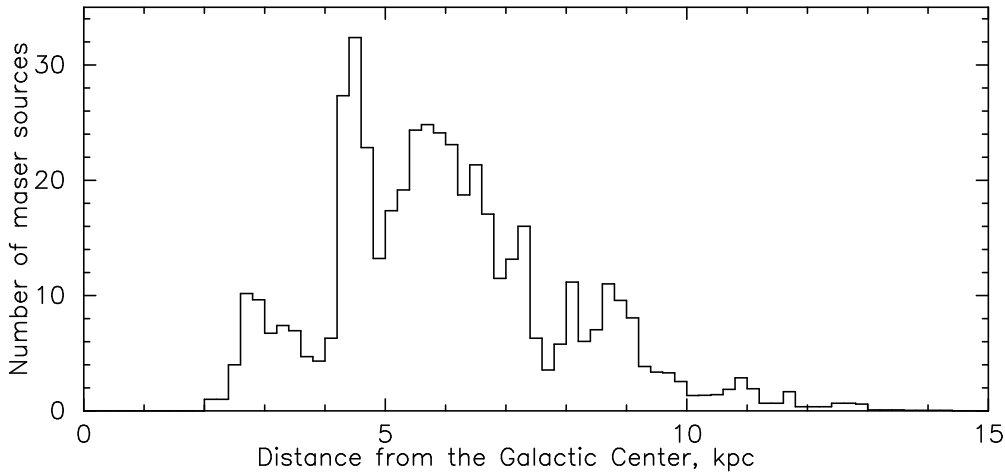}
\caption{Dependence of the maser number density on the distance
from the Galactic Center.}\label{fig:fg2}
\end{center}
\end{figure}

Guided by this hypothesis we constructed the radial dependence of
the maser number density as a function of the distance from the
Galactic Center which is shown in the figure~\ref{fig:fg2}. This
dependence shows pronounced peaks at the distances which are close
to those of the spiral arms in the current model of the Galaxy.
Maser statistics shows that the lifetime of the methanol masers is
rather limited (see, e.g.,\cite{vdw05} and \cite{szy05}). Hence,
maser sources are a good tracers of regions of the early massive
star formation which are likely to be concentrated to the spiral
arms. So, the correlation which is found does look promising for
the studies of the structure of our Galaxy.

Consideration of the selection effect related to the limited
sensitivity of the surveys leaves the maser number density peaks
at their places.

We also constructed the 6.6 GHz methanol maser luminosity function
using "near" distances. It was found that the population of masers
is greatly dominated by the low luminosity sources and future
sensitive surveys are likely to increase the number of the known
sources substantially. It should be noted that the surveys in the
regions of the spiral arm tangents are the most promising in the
sense of new detections but provide little information for the
studies of the structure of the Galaxy.

\subsection{Statistics and the maser formation regions}
Using the unified catalogue we studied the spread of the ratios of
the brightnesses in the different maser transitions as well as the
correlation between the maser ranges and the breadths of such
molecular shock tracers as CS and SiO lines (\cite{mal03}). It was
shown that: physical conditions within the usual maser source vary
considerably; maser brightness is determined by parameters of some
distinguished part of the object - maser formation region; class
II methanol masers are formed in the regions affected by the shock
propagation.

\subsection{Comparison of the methanol maser velocities with those of
CS(2-1) line}

We aimed on finding the relation between the bulk motions in the
maser formation region with those of the dense gas of the ambient
cloud. Observations show that the peak velocities of the different
maser lines often do not coincide (\cite{cas95, mal03}) and the
relative brightness of the peaks varies with time (\cite{goe04}).
In this situation the mean velocity of the maser formation region
is better represented by the center of the range of velocities of
the maser features. This should be true for the case when the
source is strong enough.

In order to get the data on the CS(2-1) velocities of the
brightest masers we carried out observations of 31 source using
the Mopra antenna. This brought us 18 new detections and a
collection of the better quality spectra. As a result we obtained
accurate CS(2-1) velocities for practically all masers with the
6.6 GHz flux density exceeding 150 Jy.

\begin{figure}
\vspace{-0.3cm}
\begin{center}
\includegraphics[width=3in,angle=0]{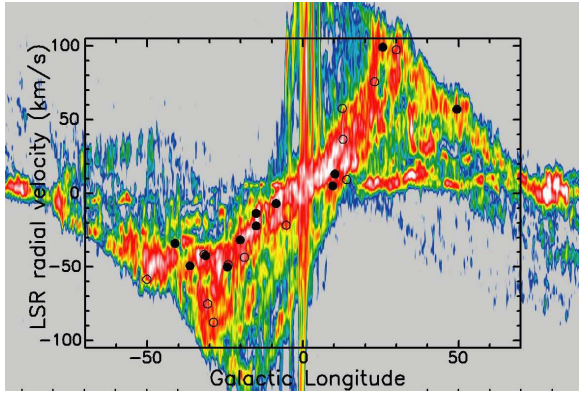}
\caption{Blue-shifted (filled circles) and red-shifted (open
circles) masers in the l-V$_{lsr}$ diagram with CO emission in the
background.}\label{fig:fg3}
\end{center}
\end{figure}

We assumed that the maser has significantly shifted velocity if
the center of the maser range was apart from the CS(2-1) line
center by more than 0.75~km/s. It was found that the difference in
the numbers of the red-shifted and blue-shifted masers is
statistically insignificant. However, the distributions of maser
sources with the different sense of the shift display considerable
difference. Figure~\ref{fig:fg3} shows that the blue-shifted maser
sources with galactic longitudes from 320$^o$ to 350$^o$ form a
distinguished group which is located in the region of the
Scutum-Centaurus spiral arm. This can reflect existence of a grand
design, i.e., grouping of the sources with similar peculiarity of
morphology or of the evolutionary stage of the massive star
forming regions.

\section{Conclusions}\label{sec:concl}

We have shown that the methanol masers represent a powerful
diagnostic tool for the studies of the nature of the massive star
formation regions and their distribution in the Galaxy.

\begin{acknowledgments}
Russian coauthors are supported by the Ministry of education and
science and RFBR under grant number 03-02-16433.
\end{acknowledgments}

\end{document}